\documentclass[aps,showpacs,amsmath,amssymb,pra,floatfix,superscriptaddress,preprint]{revtex4-1}

\usepackage{graphicx}
\usepackage{braket}
\usepackage{color}
\usepackage{dsfont}

\newcommand{\rref}[1]{Ref.~\cite{#1}}
\newcommand{\fref}[1]{Fig.~\ref{#1}}

\newcommand{\bref}[1]{(\ref{#1})}
\newcommand{\eref}[1]{Eq.~(\ref{#1})}

\newcommand{\sref}[1]{Section \ref{#1}}

\definecolor{orange}{rgb}{0.8,0.4,0.0}

\begin{document}

\title{Non-Markovian Dynamics in Ultracold Rydberg Aggregates}

\author{M. Genkin}
\affiliation{Max-Planck-Institute for the Physics of Complex Systems, 01187 Dresden, Germany}

\author{D. W. Sch{\"o}nleber}
\affiliation{Max-Planck-Institute for the Physics of Complex Systems, 01187 Dresden, Germany}

\author{S. W{\"u}ster}
\affiliation{Max-Planck-Institute for the Physics of Complex Systems, 01187 Dresden, Germany}
\affiliation{Department of Physics, Bilkent University, Ankara 06800, Turkey}

\author{A. Eisfeld}
\affiliation{Max-Planck-Institute for the Physics of Complex Systems, 01187 Dresden, Germany}

\date{\today}

\begin{abstract}
We propose a setup of an open quantum system in which the environment can be tuned such that either Markovian or non-Markovian system dynamics can be achieved. The implementation uses ultracold Rydberg atoms, relying on their strong long-range interactions. Our suggestion extends the features available for quantum simulators of molecular systems employing Rydberg aggregates and presents a new test bench for fundamental studies of the classification of system-environment interactions and the resulting system dynamics in open quantum systems.
\end{abstract}

\maketitle

\section{Introduction}

The formalism of open quantum systems, i.e., quantum systems interacting with an environment, is a widely used concept in many areas of physics. Its backbone is the separation of a large quantum system into a small system of interest and an environment, encapsulating all other degrees of freedom present in the full system. Sometimes, such an approach makes it possible to derive a tractable and physically meaningful equation of motion for the small system, rather than propagating the full system in time.
This concept~\cite{may2011,Breu_02} has become a common tool in atomic, molecular, and condensed matter systems, and also finds applications in nuclear~\cite{Anto_94,Diaz_08} and particle~\cite{Caba_05,Bert_06} physics. It is further crucially important in the field of quantum information and computation, making it possible to assess the role of decoherence in quantum information protocols~\cite{Niel_00}.

In many physical systems, the environment consists of a large number of degrees of freedom at finite temperature. Often, such an environment exhibits a back-action onto the system, which depends on previous system dynamics. In open quantum system terms, this memory of the environment is related to the concept of (non-)Markovianity.

From a practical point of view, a memoryless (Markovian) environment enables one to derive simple equations of motion, such as the Lindblad form \cite{lindblad1976}, that allow for an efficient numerical solution of the dynamics restricted to the small system space.
For strongly coupled environments with memory, typically sophisticated and numerically expensive methods are required. From this point of view it would be advantageous to possess so called quantum simulators \cite{feynman1982,feynman1986} that can capture such non-Markovian dynamics. Over the last years, several setups have been suggested with which such non-Markovian quantum simulators could be realized \cite{georgescu2014,herrera2011,mostame2012,eisfeld2012,stojanovic2012,Chiu_12,mei2013,jin2015,man2015,brito2015}.

In the present work, we propose an experimentally feasible setup where Markovian and non-Markovian dynamics can be studied in a controlled fashion using ultracold Rydberg atoms. The idea relies on the combination of two achievements, which have been reached separately in two recent experiments: Coherent oscillations of a Rydberg dimer due to resonant dipole-dipole interactions~\cite{Rave_14} and imaging of a Rydberg excitation by destroying the resonance condition of electromagnetically induced transparency (EIT) for a background gas through van-der-Waals interactions~\cite{Gunt_12,Gunt_13}. Interfacing a coupled Rydberg dimer with an optically driven background gas atom provides, on the one hand, a test bench to study the Markovian to non-Markovian transition, and on the other hand it might be useful in view of recent proposals to use Rydberg ensembles as quantum simulators for open quantum systems~\cite{Hagu_12,Scho_15,schempp2015}.

We first describe the setup in detail in \sref{sec:sec2}, and present our numerical results with experimentally accessible parameters in \sref{sec:sec3}. In \sref{sec:sec4}, the results are summarized and their implications for future work are discussed. We set $\hbar=1$ throughout the manuscript.

\section{Setup}\label{sec:sec2}

\begin{figure}[tb]
\centering
\includegraphics[width=\columnwidth]{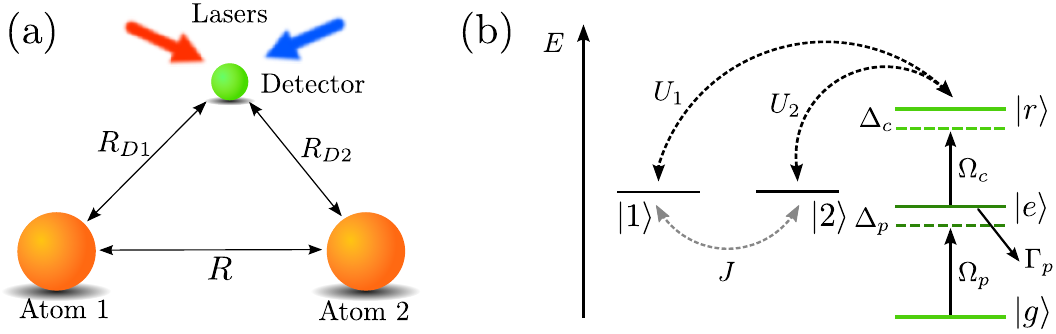}
\caption{(Color online) Sketch of the setup. (a) Atoms 1 and 2 form the Rydberg dimer with interatomic separation $R$, and the laser-driven detector atom placed in their vicinity. The distances of the detector to the dimer atoms are denoted by $R_{D1}$ and $R_{D2}$, respectively. (b) Level sketch of the setup. The dimer states $\ket{1}$ and $\ket{2}$ are coupled to each other via resonant dipole-dipole interaction with strength $J$ and interact with the Rydberg level $\ket{r}$ of the detector atom via the interactions $U_1$, $U_2$. The ground state $\ket{g}$ of the detector is coupled to the state $\ket{e}$ by the probe field with Rabi frequency $\Omega_p$ and detuning $\Delta_p$, and the state $\ket{e}$ to the Rydberg level $\ket{r}$ by the control field ($\Omega_c$, $\Delta_c$). $\Gamma_p$ is the spontaneous decay rate of the level $\ket{e}$.}
\label{fig:figsetup}
\end{figure}

The basic setup and the relevant states are sketched in \fref{fig:figsetup}. We consider two Rydberg atoms (`Rydberg dimer') in states $\ket{\alpha}=\ket{\nu\ell}$ and $\ket{\beta}=\ket{\nu'\ell'}$ respectively, with $\nu$, $\nu'$ denoting the (large) principal quantum numbers and $\ell$, $\ell'$ the angular momentum quantum numbers. The state configuration is chosen such that coherent Rabi oscillations due to resonant dipole-dipole interactions~\cite{Rave_14} are enabled between the pair states $\ket{1}=\ket{\alpha,\beta}$ and $\ket{2}=\ket{\beta,\alpha}$. The essential dynamics for the pair states specified below is thus captured in a two-state picture with the Hamiltonian
\begin{equation}\label{eq:Hsys}
H_S=J(\ket{2}\bra{1}+\ket{1}\bra{2}).
\end{equation}
Here, $J$ denotes the resonant dipole-dipole matrix element given by $J=C_3/R^3$, where $C_3$ is a state-dependent interaction coefficient and $R$ the interatomic separation of the dimer [cf. \fref{fig:figsetup}(a)]. The dimer constitutes our system $S$. We now bring a third, laser-driven atom into the vicinity of the dimer. This driven atom constitutes our environment and is from now on referred to as the detector \cite{Scho_15}. The laser field (probe field) couples the ground state $\ket{g}$ of the detector atom to some intermediate level $\ket{e}$, which in turn is coupled to a Rydberg state $\ket{r}$ by a second laser field (control field). In the rotating wave approximation, the detector is described by the Hamiltonian
\begin{equation}
H_D=\Big(\frac{\Omega_p}{2}\ket{e}\bra{g}+\frac{\Omega_c}{2}\ket{r}\bra{e}+{\rm H.c.}\Big)-\Delta_p\ket{e}\bra{e}-(\Delta_p+\Delta_c)\ket{r}\bra{r},
\end{equation}
where $\Omega_p$, $\Omega_c$ denote the Rabi frequencies and $\Delta_p$, $\Delta_c$ the detunings of the probe- and coupling fields. As Rydberg states have a very long (though finite) lifetime \cite{gallagher2005,beterov2009}, we neglect the spontaneous decay of the state $\ket{r}$ in our scheme. The intermediate state $\ket{e}$, however, is chosen to undergo radiative decay, which takes place on the time scale of the dynamics of the system. In order to account for this effect, we model the spontaneous decay with rate $\Gamma_p$ from this level by the Lindblad operator 
\begin{equation}\label{eq:lindblad}
 L=\sqrt{\Gamma_p}\ket{g}\bra{e}.
\end{equation}

In the absence of interactions between the dimer (system) and the detector (environment), the dimer dynamics is simply governed by the unitary von-Neumann equation
\begin{equation}
\dot{\rho}_S=-i[H_S,\rho_S],
\end{equation}
and the dynamics of the detector (environment) by a master equation in Lindblad form,
\begin{equation}\label{eq:EOMEIT}
\dot{\rho}_D=-i[H_D,\rho_D]-\frac{1}{2}\left(\rho_D L^{\dagger}L + L^{\dagger}L\rho_D-2 L\rho_D L^{\dagger}\right).
\end{equation}
Here, $\rho_S$ and $\rho_D$ are the density operators of system and detector, respectively. The system-environment coupling emerges due to strong van-der-Waals-type interactions between the Rydberg state of the detector with the Rydberg states of the dimer. 

Our exploitation of a single three-level atom as an ``environment'' may seem unusual, given the more typical situation where the environment is characterized by a particularly large number of quantum states. It makes sense though, since the Lindblad treatment of spontaneous decay \bref{eq:lindblad} embodies the coupling of this atom to the radiation field, which even if in the vacuum has a large number of quantum states available.

We now specify the states of the Rydberg atoms of our proposal. As in \rref{Scho_15} we take the dimer states to be $\ket{1}=\ket{ps}$ and $\ket{2}=\ket{sp}$, with $\ket{p}=\ket{43p}$ and $\ket{s}=\ket{43s}$ of ${}^{87}$Rb. These dimer states are coupled via dipole-dipole interaction, which results in a Hamiltonian of the form of \eref{eq:Hsys}, with $C_3/2\pi=1619$~MHz $\mu\mathrm{m}^3$. For the states of the detector we take $\ket{r}=\ket{38s}$, $\ket{e}=\ket{5p}$ and $\ket{g}=\ket{5s}$ \cite{Gunt_13}. Then, the interactions between the dimer states $\ket{1}$ and $\ket{2}$ and the Rydberg state of the detector are given by
\begin{subequations}\label{eq:sysenvint}
\begin{eqnarray}
U_1&=&\frac{C_4^{rp}}{R^4_{D1}}+\frac{C_6^{rs}}{R^6_{D2}},\\
U_2&=&\frac{C_6^{rs}}{R^6_{D1}}+\frac{C_4^{rp}}{R^4_{D2}}.
\end{eqnarray}
\end{subequations}
Here, $C_6^{rs}/2\pi=-87$~MHz $\mu\mathrm{m}^6$ and $C_4^{rp}/2\pi=-1032$~MHz $\mu\mathrm{m}^4$ are the interaction coefficients between $\ket{r}$ and the states $\ket{s}$ and $\ket{p}$, respectively, and the distances $R_{D1}$, $R_{D2}$ denote the separation of the detector from atom 1 and atom 2 of the dimer. The system-environment interactions \bref{eq:sysenvint} conserve the system population.
Note that our proposal does not rely on the specific states chosen, but on the state-dependence of interactions between dimer and detector, which, in principle, can also be achieved with different choices.

The system-environment interaction Hamiltonian can then be written as
\begin{equation}
H_{SD}=U_1\,\ket{1}\bra{1}\otimes\ket{r}\bra{r}+U_2\,\ket{2}\bra{2}\otimes\ket{r}\bra{r},
\end{equation}
and the master equation encapsulating the system, the environment and their interaction reads as
\begin{equation}\label{eq:fullEq}
\dot{\rho}=-i[H,\rho]-\frac{1}{2}\left(\rho K^{\dagger}K + K^{\dagger}K\rho - 2K\rho K^{\dagger}\right).
\end{equation}
Here, $\rho$ is the full density operator, $H$ the full Hamiltonian
\begin{equation}
H=H_S\otimes\mathds{1}_D+\mathds{1}_S\otimes H_D+H_{SD},
\end{equation}
$\mathds{1}$ the unity operator in a given Hilbert space and $K$ is the extension of the Lindblad operator $L$ in the full Hilbert space, $K=\mathds{1}_S\otimes L$ with $L$ given in \eref{eq:lindblad}.

\section{Numerical results}\label{sec:sec3}

In this section, we show illustrative calculations that demonstrate that, despite its simplicity, the environment provided by the detector atom is highly tunable, and in particular that the time evolution of the dimer can be tuned from Markovian to various degrees of non-Markovian dynamics. Over the last few years, a suitable measure to quantify non-Markovianity in an open quantum system has been actively pursued and debated (see e.g.\ Refs.~\cite{Riva_14,Breu_12,Breu_09,Riva_10,Haik_11,Chru_11,Luo_12,Haik_12,Rosa_12,Lore_13,Smir_13,Hall_14,Ma_14,Fanc_14,Chru_14,He_14,Addi_14,Hase_14,Hou_15,overbeck2016}), as well as used to gain insight into the dynamics of physical systems~\cite{Liu_11,Tang_12,Chiu_12,luoma2014}.
In what follows, we adopt the measure related to the information flow from the environment to the system~\cite{Breu_09}.
By this definition, the dynamics is non-Markovian whenever the trace distance between two initial density operators of the system increases at some point during their time propagation. The trace distance between two density matrices $P,Q$ is defined as
\begin{equation}\label{eq:tracedistancedef}
D(P,Q)=\frac{1}{2}{\rm Tr}|P-Q|,
\end{equation}
with $|A|=\sqrt{A^{\dagger}A}$. For a two-level system ($\ket{1},\ket{2}$), this expression simplifies to \cite{Breu_12}
\begin{equation}\label{eq:distance2level}
D(P,Q)=\sqrt{(P_{11}-Q_{11})^2+|P_{21}-Q_{21}|^2}.
\end{equation}
The rate of change of the trace distance for some initial states $P(0),Q(0)$ is given by
\begin{equation}\label{eq:sigmareduced}
\sigma(t,P(0),Q(0))=\frac{\mathrm{d}}{\mathrm{d}t}D(P(t),Q(t))
\end{equation}
and $\sigma>0$ signifies non-Markovianity. To quantify the strength of non-Markovianity given the initial states $P(0),Q(0)$, the above expression is to be integrated over all time intervals in which it takes a positive value:
\begin{equation}\label{eq:NMmeasure}
\mathcal{N}_{P,Q}=\int_{\sigma>0}dt\,\sigma(t,P(0),Q(0)).
\end{equation}
Note that to obtain an actual measure, maximization over all pairs $(P(0),Q(0))$ has to be performed in \eref{eq:NMmeasure} \cite{Breu_09,Breu_12}.  
In the following, we take initial states $\rho_1(0)=\ket{1}\bra{1}\otimes\ket{g}\bra{g}$ and $\rho_2(0)=\ket{2}\bra{2}\otimes\ket{g}\bra{g}$, which can be easily prepared (and probed) experimentally and have numerically shown to yield large values $\mathcal{N}_{\rho_1,\rho_2}$ \cite{li2010}. The corresponding system states $\rho_{S,i}(0)={\rm Tr}_{D}\rho_i(0),\,(i=1,2)$ have maximal initial trace distance $D(\rho_{S,1}(0),\rho_{S,2}(0))=1$. We propagate both states in time according to \eref{eq:fullEq} and thereupon obtain the trace distance $D_S$ and the rate $\sigma_S$ in the subsystem of interest (dimer) by tracing out the environment first and subsequently applying the definitions \bref{eq:tracedistancedef} and \bref{eq:sigmareduced}.
\begin{figure}[tb]
\centering
\includegraphics[width=\columnwidth]{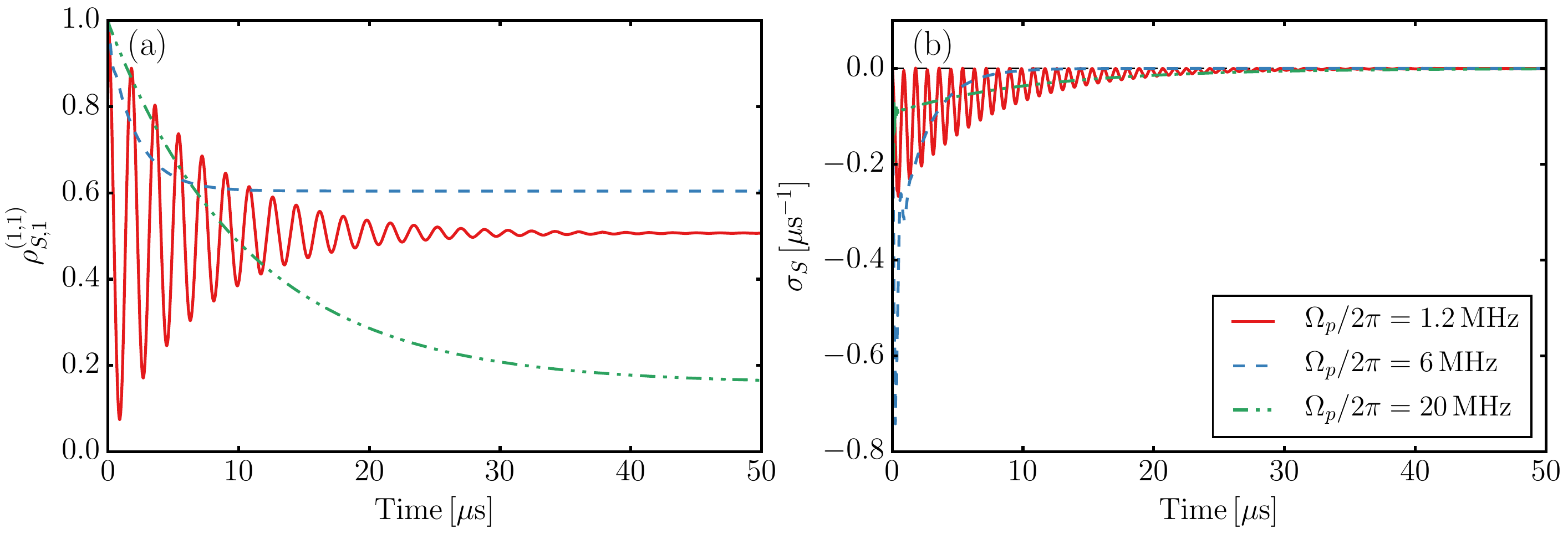}
\caption{(Color online) Dynamics of the system (dimer) for three different values of the Rabi frequency $\Omega_p$. Panel (a) shows the population of the state $\ket{1}$ for the initial state $\rho_1(0)$, and panel (b) the trace distance change rate $\sigma_S$ between $\rho_{S,1}(t)$ and $\rho_{S,2}(t)$ in the system, if system plus environment are prepared in $\rho_1(0)$ and $\rho_2(0)$, respectively (see main text). The parameters are $\Gamma_p/2\pi=6.1$~MHz, $J/2\pi=0.28$~MHz, $\Omega_c/2\pi=20$~MHz, $U_1/2\pi=-26.4$~MHz and $U_2/2\pi=-0.37$~MHz, corresponding to the interatomic distances $R=18\mu$m, $R_{1D}=2.5\mu$m and $R_{2D}=15.5\mu$m. The detunings $\Delta_p$, $\Delta_c$ are set to zero. The Rabi frequencies are $\Omega_p/2\pi=1.2$~MHz (red solid curve), $\Omega_p/2\pi=6$~MHz (blue dashed curve) and $\Omega_p/2\pi=20$~MHz (green dashed-dotted curve). As evident from the time evolution of $\sigma_S$, the three sets correspond to completely Markovian system dynamics according to the definition \eref{eq:NMmeasure}, although the population dynamics in the system shows very different equilibration time scales as well as steady-state values.}
\label{fig:figM_comparison}
\end{figure}

Before discussing non-Markovianity we illustrate how the dimer dynamics depends on the properties of the environment constituted by the detector atom, and how these properties can be tuned. In \fref{fig:figM_comparison}(a) we show different dimer dynamics arising for different Rabi frequencies $\Omega_p$ of the probe field driving the detector atom, indicating that both dephasing strength and steady-state value of the dimer dynamics can be easily controlled via the parameters of lasers acting on the detector atom.

The different strengths of dephasing can be understood on grounds of the strong asymmetry in the interactions $U_1\gg U_2$. In this way, the environment can distinguish whether the system is in state $\ket{1}$ or $\ket{2}$ and acts as a measurement device, causing dephasing and decoherence in the system \cite{Scho_15}. Consider the case when the laser fields are applied resonantly, $\Delta_p=\Delta_c=0$. The detector is then tuned to the condition of electromagnetically induced transparency (EIT)~\cite{Flei_05}, giving rise to a  so called dark state which has no contribution from state $\ket{e}$. If the dimer is in the state $\ket{2}$, the detector remains in the dark state since the interaction $U_2$ is negligible by design of the experiment.
However, if the dimer is in the state $\ket{1}$, the strong interaction $U_1$ shifts the Rydberg level of the detector $\ket{r}$ out of resonance, disturbing the EIT condition, which yields a non-zero population of the state $\ket{e}$. This state then decays with the rate $\Gamma_p$, and the emitted photons provide a potential observer with information about the state of the dimer. The stronger the driving $\Omega_p$, the more photons will be scattered by the detector atom, allowing to infer the state of the dimer more quickly, and thereby dephasing the dimer dynamics more quickly.

\begin{figure}[tb]
\centering
\includegraphics[width=\columnwidth]{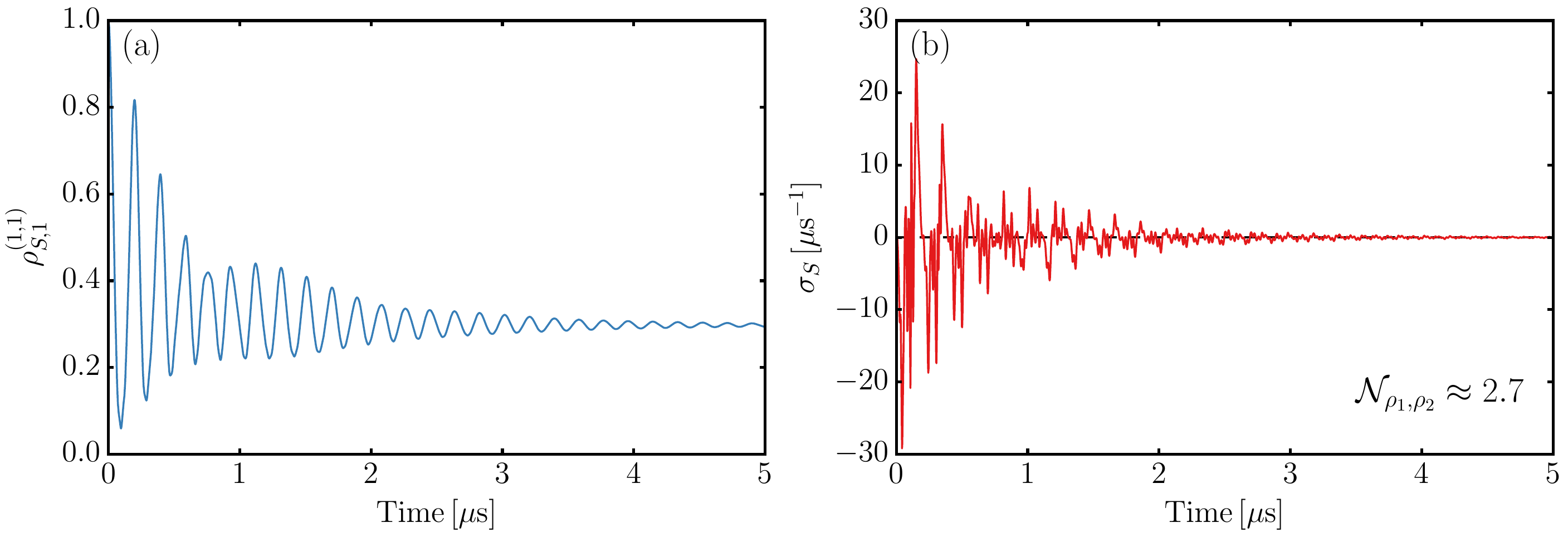}
\caption{(Color online) Same as in \fref{fig:figM_comparison} but using the parameters $J/2\pi=3.16$~MHz, $\Omega_p/2\pi=\Omega_c/2\pi=30$~MHz, $U_1/2\pi=-36.9$~MHz, and $U_2/2\pi=-0.8$~MHz, corresponding to the interatomic distances $R=8\mu$m, $R_{1D}=2.3\mu$m and $R_{2D}=8.3\mu$m. The detunings are $\Delta_c/2\pi=-\Delta_p/2\pi=50$~MHz. As evident from the time evolution of $\sigma_S$, the system dynamics is non-Markovian ($\mathcal{N}_{\rho_1,\rho_2}\approx 2.7$), also reflected in the population revival at $\approx 1\,\mu$s seen in panel (a).}
\label{fig:figNM_visible}
\end{figure}
However, as depicted in \fref{fig:figM_comparison}(b), various dimer dynamics with vastly different dephasing time scales and steady-state values can still be purely Markovian according to \eref{eq:NMmeasure}. This cautions one that looking at the population dynamics alone can be misleading when trying to estimate the Markovianity of the dynamics.

We now demonstrate the tunability of our setup. By modifying the interatomic distances as well as the laser parameters, we can switch the dimer dynamics from Markovian to non-Markovian, as shown in \fref{fig:figNM_visible}. Now, strong oscillations with $\sigma_S>0$ can be seen in \fref{fig:figNM_visible}(b), leading to a clearly nonzero $\mathcal{N}_{\rho_1,\rho_2}\approx2.7$ quantifying non-Markovianity. In the chosen configuration, the non-Markovianity of the system dynamics is not only reflected in the trace distance change rate $\sigma_S$, but can also be seen in the population dynamics \fref{fig:figNM_visible}(a) which displays a clear revival at $\approx 1\,\mu$s of the damped population oscillations.

\begin{figure}[tb]
\centering
\includegraphics[width=\columnwidth]{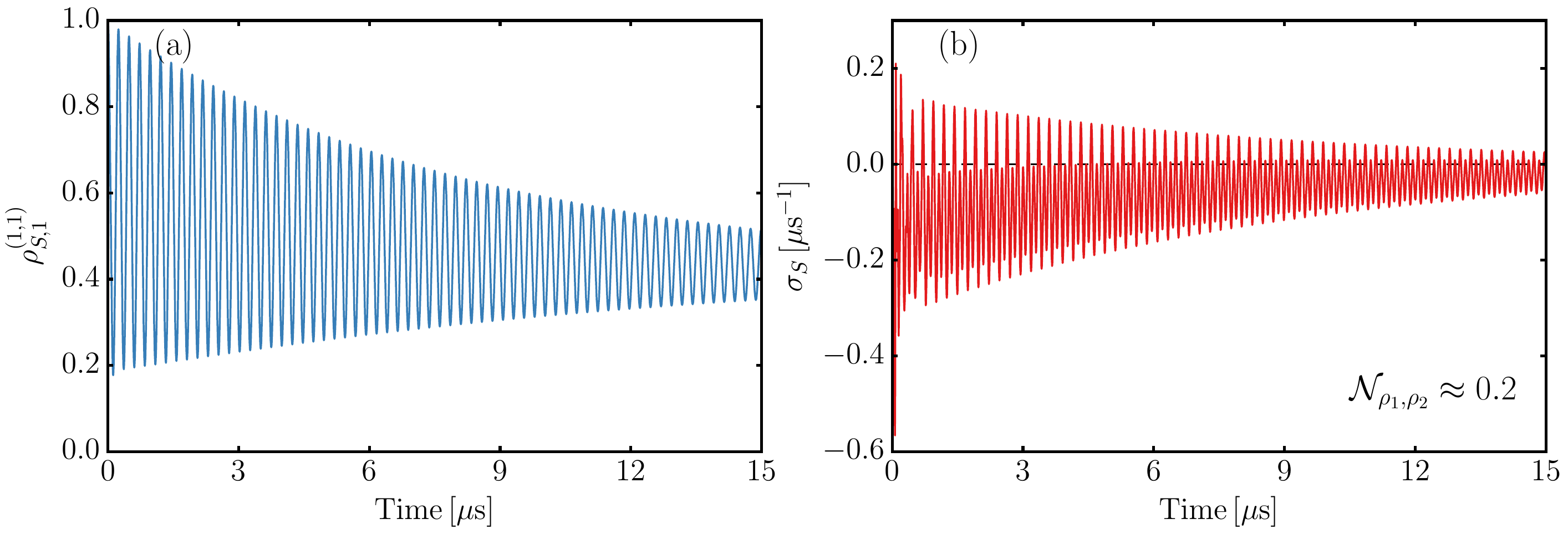}
\caption{(Color online) Same as in \fref{fig:figM_comparison} but using the parameters $J/2\pi=1.89$~MHz, $\Omega_p/2\pi=\Omega_c/2\pi=30$~MHz, $U_1/2\pi=-4$~MHz, and $U_2/2\pi=-0.11$~MHz, corresponding to the interatomic distances $R=9.5\mu$m, $R_{1D}=4\mu$m and $R_{2D}=10.3\mu$m. The detunings are $\Delta_c/2\pi=-\Delta_p/2\pi=20$~MHz. As can be seen from the time evolution of $\sigma_S$, the system dynamics is non-Markovian ($\mathcal{N}_{\rho_1,\rho_2}\approx 0.2$), however, this is not obvious from the population dynamics shown in panel (a).}
\label{fig:figNM_notvisible}
\end{figure}
It has to be noted, though, that visible non-Markovian features in the population dynamics are not necessarily present even if the system dynamics is non-Markovian. Indeed, in \fref{fig:figNM_notvisible} we show another example of non-Markovian system dynamics, in which the clearly positive contributions $\sigma_S>0$ in panel (b) lead to a nonzero $\mathcal{N}_{\rho_1,\rho_2}\approx 0.2$ while the population dynamics displayed in panel (a) does not exhibit noticeable revivals or other features often associated with non-Markovian dynamics. Comparing the figures obtained from \eref{eq:NMmeasure}, we see that $\mathcal{N}_{\rho_1,\rho_2}$ and thus the degree of non-Markovianity is significantly larger in \fref{fig:figNM_visible} than in \fref{fig:figNM_notvisible}, explaining the lack of non-Markovian features observed in the population dynamics in \fref{fig:figNM_notvisible}. Upon decreasing the rate of dissipation in the environment (spontaneous decay rate $\Gamma_p$), however, even in this setting revivals become visible.

In summary, to observe non-Markovianity in the system dynamics we have found that one needs several ingredients: 
(i)~Long detector equilibration time and intrinsic dynamics in the detector atom. Long detector equilibration time can be achieved by e.g. reducing the radiative decay rate $\Gamma_p$ (which is, however, experimentally impractical) or by introducing a large detuning $\Delta_p$ of the intermediate state while at the same time keeping the two-photon resonance condition $\Delta_p + \Delta_c \sim 0$. 
(ii)~Comparability of time scales of aggregate and detector dynamics. This can be most easily attained by tuning the aggregate coupling $J$, as the detector time scale results from a complex interplay of laser parameters, radiative decay and interactions.
(iii)~Correlation between aggregate dynamics and photon emission from the detector atom, i.e., ability to deduce the state of the aggregate by measuring the photons emitted by the detector atom. Though this condition is not fully separable from the previous one (ii), it can be met by ensuring a strong interaction $U_1$ between aggregate atom $1$ and detector atom and a strong asymmetry $U_1\gg U_2$ between the interactions $U_1$ and $U_2$ of the two aggregate atoms with the detector atom.
Whereas the first condition (i) guarantees the presence of environment memory, (ii) and (iii) guarantee the visibility of the environment dynamics in the system dynamics. 
This can be seen in Figs.~\ref{fig:figNM_visible} and \ref{fig:figNM_notvisible}: To reduce the degree of non-Markovianity in \fref{fig:figNM_notvisible} as compared to \fref{fig:figNM_visible}, we reduced the detuning $|\Delta_p|$, the interaction $U_1$ and the aggregate coupling $J$. Reducing the detuning $|\Delta_p|$ decreases the equilibration time of the detector dynamics, decreasing the interaction $U_1$ reduces the correlation between aggregate and detector, and reducing the aggregate coupling $J$ decreases the visibility of the back-action induced by the detector dynamics.

\section{Discussion and Summary}\label{sec:sec4}
The presented setup provides a test bench to study controllable non-Markovianity in open quantum systems. We have shown that both Markovian as well as non-Markovian system dynamics can be achieved by the driven-dissipative environment provided by the detector atom. Besides, our analysis reveals that (non-)Markovianity of the system (dimer) dynamics cannot be easily inferred from population dynamics alone, but rather a measure relying on the information provided by the full density matrix of the system has to be employed.

Our proposal represents a first step towards a non-Markovian quantum simulator harnessing ultracold Rydberg atoms and should be accessible by state-of-the-art experimental setups. In contrast to using the environment as a measurement device for the dimer dynamics \cite{Gunt_12,Gunt_13,Scho_15}, in our setup the single detector atom operates as gateway to the environment of electromagnetic field modes implicitly responsible for its spontaneous decay. The system dynamics can be extracted by different means \cite{Rave_14}. 

Having shown the variety of Markovian/non-Markovian dynamics as well as dephasing time scales and steady-state values of the system in the case of a simple setup employing a single detector atom, we expect even richer tunability of the dynamics in the case of many detector atoms. This might open up new prospects for using Rydberg aggregates as quantum simulators with a controlled environment.

\acknowledgments
We thank Shannon Whitlock and Kimmo Luoma for helpful discussions.

\end{document}